\documentclass[aip,
showpacs,floatfix
]{revtex4-1}
\usepackage{graphicx,amsmath,amssymb,latexsym,amsfonts,bbm}
\usepackage[usenames]{color}
\usepackage{epsfig}
\usepackage{mathtools}

\newcommand{\NAemb}{N_{\mathrm{emb}}^A}
\newcommand{\NBemb}{N_{\mathrm{emb}}^B}
\newcommand{\XC}{\mathrm{XC}}
\newcommand{\ket}[1]{\left| #1 \right>}
\newcommand{\bra}[1]{\left< #1 \right|}
\newcommand{\braket}[2]{\left< #1\vphantom{#2} \right|\left.\!\vphantom{#1}#2\right>}

\begin{document}\title{Embedding for bulk systems using localized atomic orbitals}
\author{F. Libisch}
\affiliation{Institute for Theoretical Physics Vienna University of Technology, 
A-1040 Vienna, Austria, EU}
\author{M. Marsman}
\affiliation{Department of Computational Materials Physics, 
University of Vienna, Sensengasse 8/12, A-1060 Wien, Austria, EU} 
\author{J. Burgd\"orfer}
\affiliation{Institute for Theoretical Physics Vienna University of Technology, 
A-1040 Vienna, Austria, EU}
\author{G. Kresse}
\affiliation{Department of Computational Materials Physics, 
University of Vienna, Sensengasse 8/12, A-1060 Wien, Austria, EU}

\date{\today}

\begin{abstract}

We present an embedding approach for semiconductors and insulators
based on orbital rotations in the space of occupied Kohn-Sham
orbitals.  We have implemented our approach in the popular VASP
software package. We demonstrate its power for defect
structures in silicon and polaron formation in titania, two
challenging cases for conventional Kohn-Sham density functional
theory.
\end{abstract}
\maketitle
\section{Introduction}
Ab-initio electronic structure theory for bulk materials has
experienced tremendous advances in many areas such as density
functional theory \cite{Kohn, BurkePerspective, ChallengesDFT,
  BurkeReview}, improved post-DFT 
\cite{Nozier,Langreth,Miyake,Fuchs,Furche,SOSEX,RPA}
and, e.g., van der Waals functionals,\cite{tkatchenko12} as well as
highly accurate quantum chemical \cite{MP1,MP2,PCC} and Monte-Carlo
approaches\cite{Alavi}. However, many problems are still out of reach
of an advanced theoretical description due to their size: the accurate
description of, for example, defect structures requires both a highly
accurate treatment of the local defect region, as well as the
treatment of a large number of atoms of the
environment\cite{libischprl}. It is often challenging for a single
method to meet both requirements. Embedding is therefore a suitable
strategy to evercome this hurdle. Its underlying idea is to treat the
local structure or, more generally, the subsystem of interest by a
high-level method while treating the environment with the help of a
numerically less demanding lower level method. Consistently combining
different electronic structure methods within the same calculation is
both the advantage and the challenge of the embedding approach
\cite{NeugebauerSelfc}.

Several embedding schemes have been proposed
\cite{Nafziger,Goodpaster1,Goodpaster2,Goodpaster3,Wesolovski1,Jacob,libischreview,Carter},
relying either on some form of a local embedding potential
$V_{\mathrm{emb}}(\mathbf{r})$ that mediates the interaction between
the subsystem referred in the following as the cluster and the
surrounding environment. More elaborate operator-based approaches
\cite{Knizia1,Manby} introduce a nonlocal embedding $\hat
V_{\mathrm{emb}}(\mathbf{r},\mathbf{r'})$.  Typically, subsystems are
treated in the presence of $V_{\mathrm{emb}}(\mathbf{r})$ [or $\hat
  V_{\mathrm{emb}}(\mathbf{r},\mathbf{r'})$] using a high-level
method, while the entire system is handled by density functional
theory (DFT). The individual subsystem densities are then added to
obtain an approximation for the total density of the entire system.
While conceptually simpler, local embedding potentials feature the
distinct disadvantage that no set of mutually orthogonal orbitals of
the entire system exists.  Consequently, evaluation of the total
energy becomes challenging: in particular the kinetic energy needs to
be approximated. Huang et al.~\cite{Carter} used an optimized
effective potential method to recover the kinetic energy given a total
electron density. Conversely, Fornace et al. presented an embedded
mean-field theory \cite{ManbyMeanfield} partitioning the one-particle
density matrix of the system based on its basis functions. A single
Hamiltonian then describes the entire system, avoiding any issues with
evaluating the kinetic energy for cluster and environment
separately. Additionally, this approach, by construction, allows for
direct charge exchange between the cluster of interest and the
environment. However, a direct extension to plane-wave basis sets used
in periodic solid state computations seems challenging. 
Goodpaster et al.~\cite{Goodpaster1,Goodpaster2,Goodpaster3} have
presented a scheme relying on projection operators to ensure mutual
orthogonality of orbitals belonging to different subsystems. In the
present article, we present an alternative strategy to generate and
maintain mutually orthogonal orbitals for the subsystems throughout
the calculation. We determine Wannier-like orbitals localized within the
cluster by performing unitary rotations within the subspace of fully
occupied Kohn-Sham orbitals while the orthogonal complement of
remaining orbitals resides within the environment \cite{Knizia1}. During the
optimization cycle for the cluster involving an advanced functional,
the environment orbitals remain frozen and thus orthogonality is
preserved. This approach avoids the inaccuracies associated with
approximating the kinetic energy.

In the present paper we demonstrate the power of our embedding scheme
in a proof-of-principle calculation adressing two problems for which
standard Kohn-Sham DFT is known to be inadequate: defects in silicon
and polarons in titania. We use the following hierarchy of methods:
the cluster is treated by the (expensive) hybrid functional PBEh while
the environment is treated only by the PBE functional. We show that
this embedding scheme implemented in the Vienna Ab Initio Simulation
Package (\texttt{VASP}) is robust and efficient. We emphasize that the
present embedding scheme is not limited to hybrid-DFT in DFT
embeddings. Future extensions will adress the treatment of the cluster
by RPA or quantum chemistry approaches.

\section{Technique}

We partition a system into two parts: a cluster of interest $A$ with
atomic sites $r_{j,A}$, ($j=1,\ldots,M_A$) with $M_A$ the number of
atomic sites included in the cluster, and the surounding environment
$B$, containing $M_B$ atomic sites $r_{j,B}$, ($j=1,\ldots,M_B$). In a
first step, the entire system ($A+B$) is solved using a single,
comparatively cheap exchange-correlation functional, e.g., PBE
\cite{PBE},
\begin{equation}
    H \ket{\psi_i} = \overline\varepsilon_i\ket{\psi_i},
\end{equation}
yielding Kohn-Sham orbitals $\ket{\psi_i}$ with orbital energies
$\overline\varepsilon_i$ and the density matrix
\begin{equation}
\ket{\psi_i},\quad \gamma(\vec r,\vec r\,') = \sum_{i=1}^{N_{\mathrm{tot}}} f_i\braket{\psi_i}{\vec
  r}\braket{\vec r\,'}{\psi_i},\quad
\end{equation}
 with occupation numbers $f_i \in [0,1]$, where the index $i =
 1,\ldots,N_{\mathrm{tot}}$ goes over all orbitals and physical
 spin. Note that we we have not included $k$-point sampling in the
 present ansatz, since it is not straightforward to treat the
 transformations at different $k$-points independently. We aim to find
 a unitary rotation within the subspace of fully occupied orbitals
 ($f_i=1$, $i = 1,\ldots,N$) that yields a set of orbitals aligned
 with the atomic orbitals $\ket{\alpha_k}$ localized around the atomic
 sites of the cluster. The index $k=1,\ldots,N_A$ of the atomic
 orbitals includes both the site index as well as radial and angular
 momentum quantum numbers. To this end we apply to the orbital overlap
 matrix W,
\begin{equation}\label{eq:W}
W_{ki} := \braket{\alpha_k}{\psi_i},\quad k=1\ldots N_A,\quad i=1\ldots N.
\end{equation}
a singular value decomposition according to
\begin{equation}
   W = U\cdot D \cdot V^\dagger
\end{equation}
with $D = \mathrm{diag}(\sigma_i)$.
The unitary matrix $V$ represents the rotation in the space of the $N$ occupied 
orbitals that optimally aligns $N_A$ of these orbitals with the
atomic orbitals $\ket{\alpha_i}$ keeping the remaining $N-N_A$ orbitals
orthogonal. The singular values $\sigma_i$ provide a measure for the degree
of overlap between the orbitals $\ket{\alpha_i}$ and the rotated orbitals $\ket{\phi_i}$,
\begin{equation} \label{eq:rotation}
V\ket{\psi_i} = \ket{\phi_i},\quad
\left|\braket{\alpha_j}{\phi_i}\right| \left\{ \begin{array}{cc} \leq
  1, & i \leq N_A \\  = 0,& i > N_A\end{array}\right.
\end{equation}
Orbitals with indices $i > N$ outside the space of occupied orbitals
are unaffected by the rotation, $\ket{\phi_i} = \ket{\psi_i},\,\forall i
: N < i < N_{\mathrm{tot}}$. Using the rotated $\ket{\phi_i}$, we can
thus partition the occupied space into orbitals which have an overlap
with the $\ket{\alpha_i}$, and those who do not. Ideally, if the Kohn
Sham orbitals are well covered by the atomic wavefunctions, we expect
the singular values $\sigma_i$ to be close to 1.

After the orbital rotations, a subset $\ket{\phi_i^A}$ of those
$i=1,\ldots,\NAemb$ with $\NAemb\le N_A$ can now be optimized based on
a more expensive exchange-correlation (XC) functional $E_{\XC}^A$,
e.g., a hybrid functional \cite{hybrid1,hybrid2}, while freezing the
orthogonal complement of $\NBemb = N-\NAemb$ environment orbitals
$\ket{\phi_i^B}$.  In general, the number of orbitals $\NAemb$ used in
the embedding procedure may be smaller than the number of atomic basis
functions $N_A$: in principle, one may choose any subset of the
localized orbitals $0 < \NAemb \le N_A$. In practice, we sort the
rotated orbitals by their singular values $\sigma_i$, and choose the
$\NAemb$ orbitals corresponding to the largest $\sigma_i$, where
$\NAemb$ is chosen according to the number of orbitals of interest
within the cluster. A typical cut-off will be $\sigma_i > 0.5$. We
find that our results do not strongly depend on $\NAemb$, as long as
the number of optimized orbitals is sufficiently large as to properly describe the
local bonding.  We will discuss the choice of $\NAemb$ in more detail
in the results section below.

Note that after the orbital rotation, the $\ket{\phi_i}$ are no longer
eigenvectors of the Kohn-Sham Hamiltonian $H$. The diagonals of the
Hamiltonian are given by the expectation values
\begin{equation}\label{eq:estart}
\varepsilon_i := \bra{\phi_i}H\ket{\phi_i},
\end{equation}
related to the original eigenvalues through the invariance of the trace
\begin{equation}
\quad \sum_{i=1}^{N} \varepsilon_i = \sum_{i=1}^{N} \overline\varepsilon_i 
\end{equation}
under unitary rotations. In the case of fractional occupations of some
of the orbitals in $N_{\mathrm{tot}}$, the above considerations remain
valid in the subspace of the $N \le N_{\mathrm{tot}}$ fully occupied
orbitals. All orbitals with fractional occupation are assigned
to cluster $A$, even if they are not well localized. However, extension
of the present approach to quantum-chemistry based correlated
wavefunction approaches may face difficulties for such delocalized
metallic states.

Taking into consideration that different exchange-correlation
functionals will be employed for the cluster $A$ and the environment
$B$ we write the energy functional for the entire system $A+B$ as
\begin{equation}\label{eq:etotal}
E = \sum_{i=1}^{N_{\mathrm{tot}}}f_i \bra{\phi_i}T\ket{\phi_i}+ \frac 12 E_H[\rho] + E_{\XC}[\gamma_A,\gamma_B] 
\end{equation}
where $\rho(r) = \gamma(r,r)$ is the density, and
$E_H[\rho]$ the Hartree energy
\begin{equation}
E_H[\rho] = \frac 12 \sum_{i,j=1}^{N_{\mathrm{tot}}}f_if_j\bra{i\;j}\mathcal{C}\ket{i\; j},
\end{equation}
with $\mathcal{C}$ denoting the Coulomb operator.
The mixed exchange-correlation functional $E^{AB}_{\XC}[\gamma_A,\gamma_B]$ containing
both the lower ($B$) and higher ($A$) level functionals can be written as
\begin{equation}\label{eq:xcab}
  E_{\XC}^{AB}[\gamma_A,\gamma_B] =  E_{XC}^A[\gamma_A] + E_{\XC}^B[\gamma_B] + E^{\mathrm{int}}_{\XC}[\gamma_A,\gamma_B].
\end{equation}
To use Eq.~(\ref{eq:xcab}) in practice within our embedding approach, the contribution due to the interaction between the two subsystems, $E_\XC^{\mathrm{int}}$, should be
approximated by the lower-level functional ($B$) applied also to the environment, i.e.,
$E_\XC^{\mathrm{int}}[\gamma_A,\gamma_B] = E_\XC^{\mathrm{int},B}[\gamma_A,\gamma_B]$ with
\begin{equation}\label{eq:excint}
E_\XC^{\mathrm{int}}[\gamma_A,\gamma_B] = E_\XC^{B}[\gamma_{\mathrm{tot}}] - E_\XC^B[\gamma_A] - E_\XC^B[\gamma_B] 
\end{equation}
This allows for very expensive functionals to be used in the cluster
$A$, including RPA or quantum chemistry approaches. The drawback is
that the error introduced in such a mixed approach is difficult to
quantify a priori.

For the special case of hybrid functionals chosen in the present work as the ``higher-level'' functional $A$ the interaction term $E_\XC^{\mathrm{int}}$ can be much more accurately approximated by the full hybrid functional itself
\begin{eqnarray}
E_{\XC}^{\mathrm{int}}[\gamma_A,\gamma_B] &=& E_{\XC}^A[\gamma_A+\gamma_B] - E_{\XC}^A[\gamma_A] - E_{\XC}^B[\rho_B] 
\end{eqnarray}
as the most expensive summation $(i,j \in B)$ is constant and hence
not relevant for the optimization of orbitals in $A$. The only
approximation here is that the orbitals in $B$ are kept frozen. Even
when optimizing large supercells, only a subset $N_A \times
N_{\mathrm{tot}}$ of the full $N_{\mathrm{tot}}\times
N_{\mathrm{tot}}$ orbital pairings needs to be calculated to evaluate
the relevant exchange contribution $\bra{i\;j}\mathcal{C}\ket{j\; i}$,
greatly reducing the numerical effort.

The orbitals in $A$ can now be efficiently optimized minimizing the
energy functional of Eq.~(\ref{eq:etotal}), while the orbitals in
$B$ are kept frozen. Consequently, during a single optimization,
any change in the electronic structure of $A$ due to a
more accurate XC functional cannot lead to a redistribution of charge
in $B$. The embedding can now be made self-consistent by alternating
between subsystems $A$ and $B$ in freeze-and-thaw cycles: after the
initial solution of the entire system using the lower-level
functional, an orbital rotation is performed to partition into orbital
sets $A$ and $B$. Then, starting with $A$, alternatingly one of the
subsystems is optimized while the other one is kept frozen. Each cycle
that optimizes the orbitals $A$ uses the higher-level XC functional
$A$ while each cycle that optimizes the orbitals in $B$ employs the
lower-level functional $B$. For the example of polarons in titania discussed below, we find rapid convergence after about six freeze-and-thaw cycles \cite{NeugebauerSelfc}.

Since the interaction of the cluster region (e.g. defect) with its
periodic image needs to be minimized, conventional defect modeling is
hampered by the requirement of large supercells. If the bulk material
could well be described by conventional XC functionals, and only the
defect structure requires more advanced techniques, conventional
techniques still require an expensive evaluation of the entire exchange
contribution. The embedding procedure outlined above is ideally suited
to significantly reduce computational effort while retaining high
accuracy.

\section{Implementation in VASP} 

We have implemented the embedding scheme outlined above in the Vienna
ab initio simulation package (\texttt{VASP}) using the projector
augmented wave method of Bl\"ochl in the implementation of Kresse and
Joubert\cite{Bloechl,Kresse,KresseFuerthau}. Usage is simple: in a first step, a
conventional DFT calculation of a system is performed. In a second
step, the localized atomic basis functions $\ket{\alpha_i}$ are
defined.  Our implementation currently supports the PAW basis
functions (the pseudo partial waves) and standard spherical harmonics
(including hybrid orbitals such as $sp^3$) with the radial dependence
taken from suitably scaled hydrogen functions. \texttt{VASP} then
starts an embedded calculation, performs the orbital rotation and
optimizes the set of $N_{\mathrm{emb}}$ orbitals localized on the
cluster $A$ while the remaining fully occupied Kohn-Sham orbitals of
the environment $B$ are frozen (for spin-polarized calculations the
two spin components are handled independently in terms of the rotation
and the number of optimized orbitals). During the freeze-and-thaw
cycles, no further localization procedure according to
Eq.~(\ref{eq:rotation}) is required, and orbital
sets $A$ and $B$ are interchanged.
To enforce orthogonality between the currently optimized orbitals and
the frozen environment we use the frozen orbitals as projector.

We do not currently support $k$-point sampling in the embedding
calculation, since the orbital rotations at different $k$-points are
not independent. Likewise, forces are not currently implemented in our formalism.
The geometries used in this work were taken from Ref.~\onlinecite{MerzukDefects} 
for the defects in silicon and where relaxed using the HSE functional
similar to Ref.~\onlinecite{PolaronCesare} for
the polarons in titania.

To compare final energies calculated from the functional
Eq.~(\ref{eq:etotal}), we need to evaluate the full exchange energy
with all orbitals once after self-consistently converging the orbitals
of $A$.  We benchmark our embedding approximation against a fully self
consistent optimization of all orbitals using the hybrid
functional. Additionally, we compare against evaluating the hybrid
functional with orbitals obtained from using a conventional functional
(PBE). Obviously, both the embedded and the PBE orbitals are, by
construction, not self-consistent with respect to the hybrid
functional. In comparison with a full self-consistent optimization, a
single evaluation step using the full hybrid functional with non-self
consistent orbitals takes a small amount of time, while substantially
improving the accuracy: errors in an inexact evaluation of the
interaction between subsystems in Eq.~(\ref{eq:excint}) are
eliminated. We denote corresponding energies by an asterisk ($^*$) in
the following.

\section{Results}
\subsection{Point defects in silicon}

\begin{table}
\begin{tabular}{ccrrrr}
\multicolumn{4}{c}{energies [eV] (error [meV])}  \\
Defect& Hybrid & \multicolumn{2}{c}{DFT$^{*}$}  &  \multicolumn{2}{c}{Embedding$^{*}$} \\
\hline
  H        &  3.00  &  2.99 &(-17)  &    3.01 &(10)\\
  X        &  3.01  &  3.01 &  (1)  &    3.04 &(27)\\
  C$_{3V}$  &  3.05  &  3.03 &(-19)  &   3.06 &(18)\\
  T       &  3.77  &  3.34 &(-423) &    3.75 &(-20)\\
 VJT      &  4.14  &  4.52 &(377) &    4.19 &(44)\\
  V       &  4.23  &  $\phantom{ii}$4.83 &(599)  &$\phantom{ii}$  4.26 &(25)\\
\hline
\end{tabular}
\caption{Comparison of defect formation energies for different defect
  types in silicon. All energies in eV, errors in brackets [meV] are
  deviation from full hybrid benchmark calculation. DFT and embedding
  calculation energies represent the ``single-shot'' evaluation of the
  full hybrid energy functional using the corresponding non-hybrid PBE
  or embedded orbitals as indicated by an asterisk.}\label{tab:resultsdefects}
\end{table}

As a first practical test of our new algorithm, we consider point
defect structures in silicon \cite{MerzukDefects}. We use a 64 atom
supercell, and neglect $k$-point sampling both in the full hybrid
benchmark and in the embedding calculations. Due to the localized
nature of the covalent bonds involved, conventional Kohn-Sham DFT
fails to correctly reproduce experimental observations. By using
hybrid functionals or even more advanced RPA formulations
\cite{MerzukDefects}, these problems are mitigated. However,
comparison with more accurate correlation functionals such as RPA and
experiment show that currently available methods yield a wide range of
predictions depending on the employed functional \cite{TkatchenkoExp},
highlighting the necessity to move towards higher level
correlated wavefunction approaches.

Due to the large
supercells required to avoid interaction of the defect sites with
their periodic images, embedding the orbitals close to the defect site
seems desirable. As benchmark for our embedding scheme, we consider
defect formation energies of a set of common interstitial defects and
vacancies. We aim to reproduce the energetics of full hybrid
functional calculations based on PBEh by a cheaper embedding
calculation in which only a few (six to ten) orbitals localized in the
immediate vicinity of the defect (taken to be the $A$ orbitals of the
cluster) are treated using the hybrid functional, while the
remaining ~118-122 orbitals (taken to be the $B$ orbitals) are only
treated by PBE. We also compare our results to purely DFT-based
predictions.

For all Si defect calculations, we choose as atomic orbitals
$\ket{\alpha_i}$ the PAW pseudo-partial waves of the Si atoms at and
directly adjacent to the defect site, resulting in $N_A=16$
(vacancies) or 20 atomic basis functions (1 $s$ + 3 $p$ per atom) for
most cases. Increasing the number of basis functions per atom
increases the overall overlap of the occupied Kohn-Sham orbitals with
the defect site at the cost of a larger $N_A$, and thus a larger
overlap matrix $W \in \mathbb{C}^{N_A\times N}$ of
Eq.~(\ref{eq:W}). Consequently, the number of singular values
$\sigma_i$ increases. To obtain a set of orbitals well
localized at the defect site, we choose all orbitals with singular
values $\sigma_i > 0.5$ as embedded orbitals. This procedure yields a
number of selected embedded orbitals $\NAemb$ from 9 (X defect) to 16
(T-defect), in line with the number of Si-Si bonds one would expect for
the respective defect sites. For example, each of the two defects
atoms of the dumpbell defect (X) interacts strongly with four close
neighbors in the surrounding lattice and with the other atom in the
dumpbell, yielding a total of nine covalent bonds. Indeed, we find
nine singular values substantially larger than 0.5 for this
defect. The PAW basis functions we choose yield a set of orbitals with
a bimodal distribution: a significant number of orbitals with
$\sigma_i \approx 1$, well seperated from delocalized orbitals with
small overlap $\sigma_i \approx 0$ with the defect site. The threshold
of 0.5 is therefore a good compromise between choosing all $N_A$
possible orbitals (which will include orbitals with very small
singular values) and too few orbitals that will not allow for
reasonable optimization. Note, however, that care must be taken to
check that there are no singular values close to 0.5, to avoid
arbitrarily including (or discarding) orbitals upon small fluctuations
in $\sigma_i$.

As mentioned in Sec.~II, calculating the defect formation energies
from the total energies of Eqs.~(\ref{eq:estart})-(\ref{eq:excint})
for the defects and the defect-free (bulk) system proves
challenging. For the embedded case, vacancies and interstitials change
the number of orbitals, and make a comparison of absolute energies
problematic. We therefore compare the predictions by different methods
for the formation energies by evaluating the same hybrid energy
functional with the help of all orbitals (i.e., not just the ones localized at
the defect) generated by the different methods [denoted by an
  asterisk (*)].

Our results for various defect structures are summarized in
Tab.~\ref{tab:resultsdefects}. Overall, we find excellent agreement
between the hybrid functional benchmark and our embedding approach.
For simple, non-metallic defects such as the dumbbell configuration
(X), the hexagonal hollow (H) and a lower-symmetry variant (C$_{3V}$)
we find that both the embedding, as well as the evaluation of the
hybrid functional with the low-level DFT orbitals produces good
agreement with benchmark calculations (see second column of
Tab.~\ref{tab:resultsdefects}). By contrast, the metallic tetragonal
site is badly described by DFT: it features one interstitial Si atom
coordinated to its four nearest neighbors, so that the local
coordination of the interstitial is identical to the other Si
atoms. This position is unique insofar that the highest occupied
orbital is threefold degenerate ($t_2$ symmetry) but only occupied by
two electrons. This degeneracy is preserved in DFT yielding three
fractionally occupied orbitals with occupation numbers $f_i =
2/3$. Consequently, the evaluation of the hybrid energy functional
based on these orbitals fails to yield reasonable formation
energies. By contrast, the embedding method locally breaks the
degeneracy as does the full hybrid calculation, leading to good
agreement of the embedding results with the benchmark (see T, VJT and V
lines in Tab.~\ref{tab:resultsdefects}). 

Our results compare poorly with experimental data: one important
reason is the interaction of periodic images of the defects to the
supercell size. We therefore consider a larger supercell of 512 atoms,
still with a single defect. We note that on our hardware, the full
hybrid calculations takes ten times as long as the embedded one, with
a relative error of 0.4\% in total energy. We
find a substantial change in results for the larger cell (compare
Tab.~\ref{tab:resultsdefectslarge}), that now fit well to experimental
results for the H defect. To achieve better agreement also for
vacancies (V) requires a more accurate treatment of electronic
correlation (e.g., RPA) or inclusion of Van der Waals contributions
\cite{TkatchenkoExp}.
  
\begin{table}
\begin{tabular}{lcrrrrc}
\multicolumn{4}{c}{energies [eV] (error [meV])} &
\\ Defect& Hybrid &
\multicolumn{2}{c}{DFT$^{*}$} & \multicolumn{2}{c}{Embedding$^{*}$} &
Experiment \\ \hline 
T & 4.97 & 5.17 &(198) & 5.13 & (165) \\ 
H & 4.22 & 4.19 &(38) & 4.25 & (28) & 4.2-4.7\\
V & 5.06 & 5.55 &(482) & 5.08& (22)& 2.1-4.0\\
\hline
\end{tabular}
\caption{Same comparison as in Tab.~\ref{tab:resultsdefects} for an eight
  times larger supercell containing 512 atoms. Rightmost column shows
the range of available experimental data taken from [\onlinecite{Exp1,Exp2,Exp3,Exp4,Exp5,Exp6,Exp7,Exp8,Exp9,Exp10}].}\label{tab:resultsdefectslarge}
\end{table}

\subsection{Polarons in titania}\label{sec:polarons}

\begin{table}
\begin{tabular}{llc|rcc}
Method&& cycles & $\;E_{\mathrm{dist}}$  [eV]& $\;E_{\mathrm{ideal}}$ [eV]& $\;\Delta E$  [meV]\\
\hline
Hybrid&  &-& -972.85   & -972.33 &  514\\
\hline
DFT&&-& -688.17   & -687.81 & -355\\
DFT$^{*}$&  &-& -962.81   & -962.54 &  270\\
\hline
\multicolumn{2}{l}{Embedding$^{*}$} & 1 & -963.70 & -963.52 & 174\\
$N_A=30$&$\NAemb=14$     & 2 & -963.36 & -963.12 & 240\\
    &     & 3 & -963.47 & -963.16 & 313\\
    &     & 4 & -963.41 & -962.98 & 426\\
    &     & 5 & -963.45 & -962.99 & 459\\
    &     & 6 & -963.44 & -962.98 & 462\\
    &     & 7 & -963.45 & -962.98 & 475\\\hline
$N_A=30$&$\NAemb=6$      &7& -963.40& -962.94& 460\\
&$\NAemb=8$      &7& -963.41& -962.95& 459\\
&$\NAemb=12$      &7& -963.45& -963.00 &454\\
&$\NAemb=20$      &7& -963.50 & -963.02& 474\\
&$\NAemb=30$      &7& -963.59 &-963.08 &514\\\hline
$N_A\!=\!90$&$\NAemb\!=\!30$      &7& -963.55 &-963.08 &462          \\
$N_A\!=\!\phantom{9}6$&$ \NAemb\!=\!\phantom{3}6$      &7& -963.38 &-962.96 &420           
\end{tabular}
\caption{Energies of the distorted $E_{\mathrm{dist}}$ and ideal
  $E_{\mathrm{ideal}}$ lattice structure of charged rutile
  titania. The energy $\Delta E$ is gained by forming a small
  polaron. Energies after seven iterations are given for different
  sizes of the embedded region. Different methods are labeled as
  follows.  Hybrid: full hybrid functional calculation used as
  benchmark; DFT: direct evaluation of the energies using the PBE
  functional; DFT$^*$: evaluation of the hybrid energy functional
  using the orbitals from the PBE calculation; embedding: embedding
  calculations as function of the number of freeze-and-thaw cycles at
  fixed number of localized atomic orbitals $N_A$. After the unitary
  rotation to localize orbitals around the defect
  [Eq.~(\ref{eq:rotation})], we optimize a subset $\NAemb \le N_A$ of
  orbitals in the embedded region $A$ as noted.}\label{tab:selfcons}
\end{table}

As a second demonstration of our method, we consider the formation of
polarons in titania. We consider a $2\times 2\times 2$ supercell with
24 Ti and 28 O atoms in the rutile structure. In an accurate hybrid
functional description, an additional electron localizes, distorting
the lattice and forming a small polaron. The distortion decreases the
energy compared to a delocalized charge. A full hybrid functional
calculation yields a decrease in energy by 514 meV for the distorted
geometry.  We will use this value in the following as benchmark for
our embedded description of the polaron. Concerning the atomic basis
functions $\ket{\alpha_i}$ used for the initial localization, we
typically use the 1 $s$ and 5 $d$ orbitals of the Ti atom centered at
the small polaron deformation, as well as all $s$ and $p$ orbitals of
the six nearest neighbor oxygen atoms, yielding a total of $N_A=30$
orbitals.  Geometries for the distorted structure were relaxed using
HSE calculations.

\begin{figure*}
  \includegraphics[width=\textwidth]{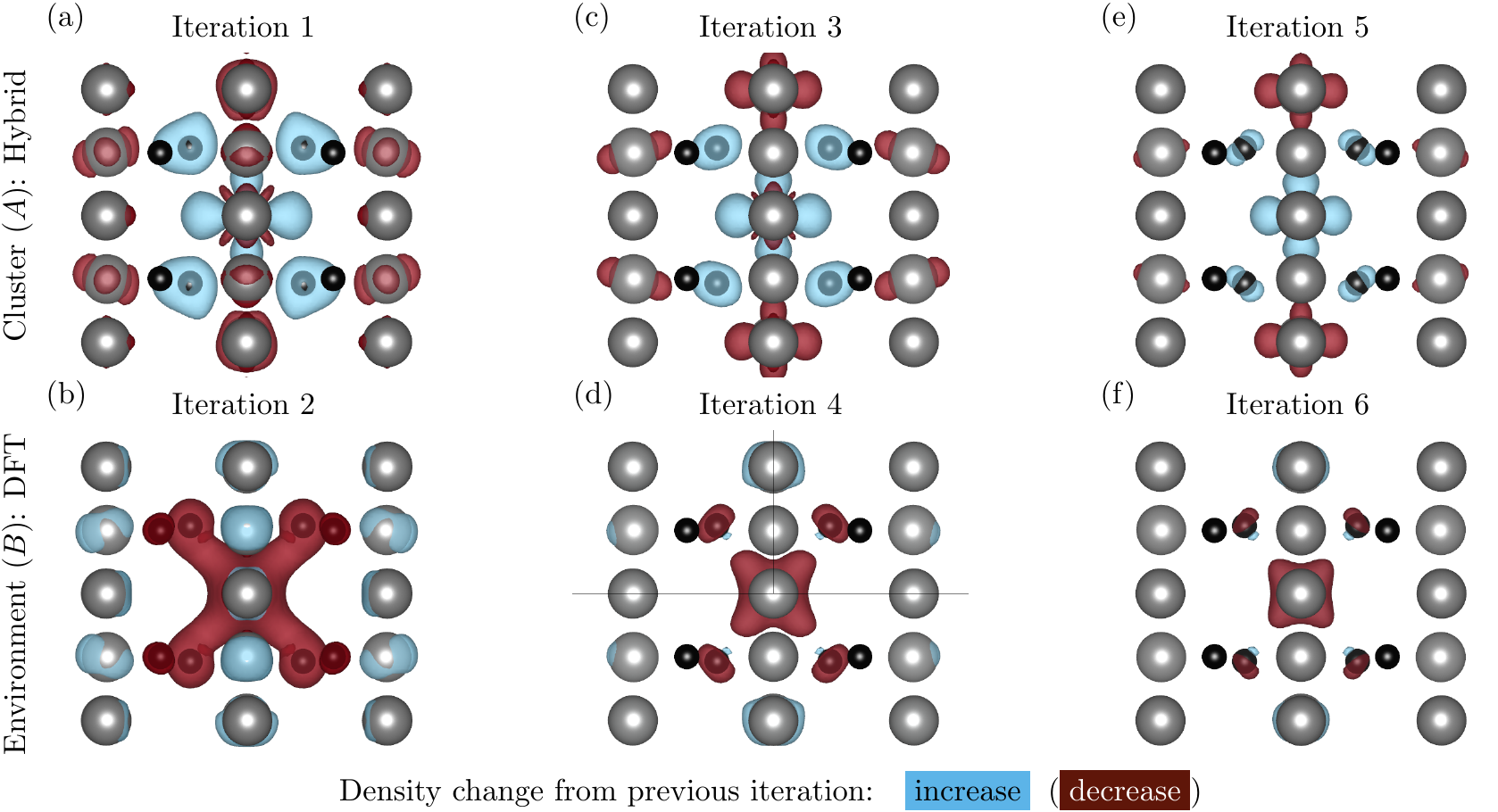}
\caption{Isosurface plot of charge variation $\Delta\rho$ in the
  density compared to the previous iteration (for iteration 1 the
  difference to DFT is shown).  Yellow (teal) denotes density
  decrease (increase). Odd iteration numbers (top row) correspond to
  optimization in the cluster $A$ using the hybrid functional, while
  even iteration numbers (bottom row) correspond to PBE optimizations
  of the environment $B$.}\label{fig:convergence}
\end{figure*}

Our results for the small polaron formation energy are summarized in
Tab.~\ref{tab:selfcons}. Conventional density functional theory
invoking a PBE functional is not capable of reproducing small polaron
formation, predicting even a negative energy gain (i.e. energy costs)
of -355 meV to form the polaron. Inserting the DFT orbitals in the
hybrid energy functional leads to a correction of the sign. However
the energy gain is underestimated by a factor of two (270 meV), see
Tab.~\ref{tab:selfcons}. A single-cycle embedding calculation yields a
slightly larger error predicting 190 meV. The origin of this error is
obvious: while the hybrid functional tries to localize the charge in
the cluster region $A$, the surrounding region $B$ cannot react to the
substantial change in the electrostatics, since all $B$ orbitals are
frozen.

Subsequent freeze-and-thaw cycles rapidly improve the result: we
alternate between optimizing the two sets of orbitals $A$ and $B$, one
with the expensive hybrid, the other with pure GGA (PBE). We find
convergence in about seven iterations, quite independent of the number
of embedded orbitals $\NAemb$ (Tab.~\ref{tab:selfcons}). As minimum
requirement for $\NAemb$, the Ti atom at the center of the distortion
and the surrounding oxygen atoms need to be treated accurately, which
is already achieved with as few as six orbitals
(Tab.~\ref{tab:selfcons}). Note that only choosing the central Ti atom
as atomic basis, $N_A = \NAemb = 6$, yields a smaller polaron energy
than chosing the six orbitals with the highest singular values from
the $N_A=30$ localized orbitals including also the closest oxygen
atoms. The reason is that in the latter case, the response of the
surrounding shell of oxygen atoms is - to some degree - also treated
by the hybrid functional. However, further increasing the number of
localized orbitals $N_A$ by, e.g., also including a shell of
neighboring Ti atoms does not result in a stronger overlap of $A$
orbitals on the central Ti atom (note that the localization
procedure does not distinguish between the different atomic basis
functions $\ket{\alpha_i}$). Consequently, such a large $N_A$ would
require a comparatively large $\NAemb$ to ensure that orbitals close
to the central site are included in the embedded
calculation. Otherwise accuracy is lost. Indeed, we find a better
agreement with the benchmark for $N_A = \NAemb = 30$ than for
$N_A=90$, $\NAemb = 30$ (see Tab.~\ref{tab:selfcons}).  Since the
numerical effort of the embedding calculation scales with a power of
$\NAemb$, in practice a small $N_A$ that allows for $N_A \ll N$ is
preferable.

\begin{figure*}
\includegraphics[width=\textwidth]{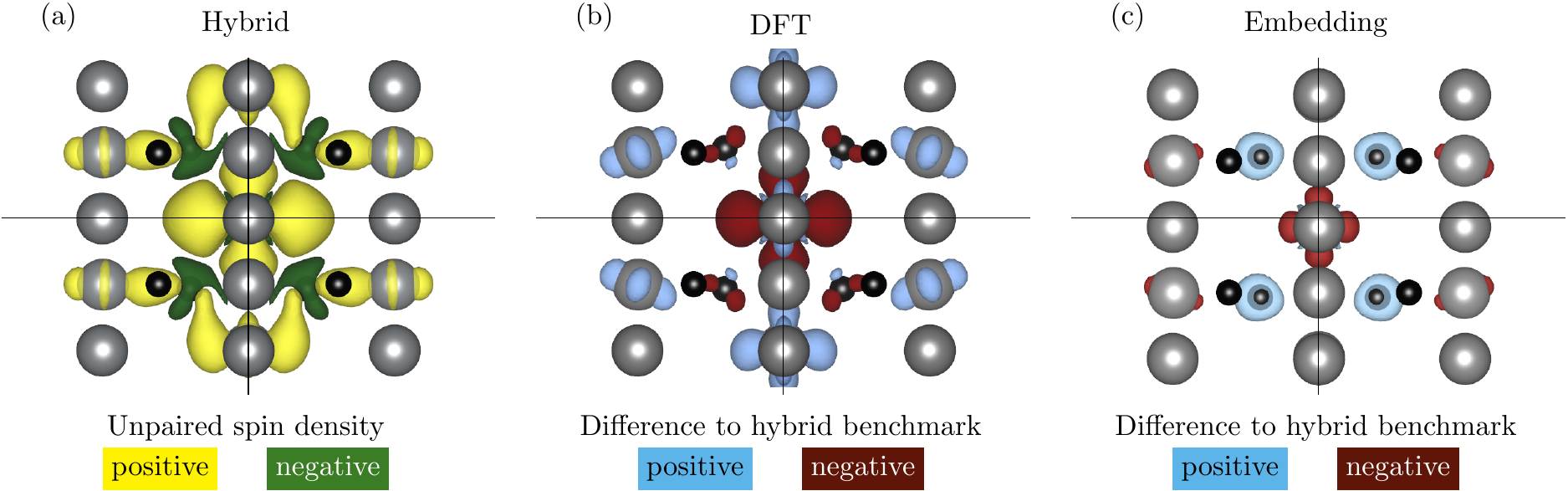}
\caption{(a) Isosurface plot of the spin density ($\rho_+ - \rho_-$)
  of the converged polaron wavefunction in titania, using a full
  hybrid functional calculation, seen from the (100) direction. Blue
  (red) spheres correspond to Ti (O) atoms. The image is centered
  around the Ti atom at the center of the distortion. Tourquois
  (yellow) denotes positive (negative) signs. (b,c) Isosurface plot of
  the difference in unpaired spin density between (b) DFT [(c)
    embedding] and the hybrid benchmark (a). Blue (orange) corresponds
  to a density increase (decrease) compared to
  (a).}\label{fig:methodcomp}
\end{figure*}

It is intstructive to follow the charge density variations along the
freeze-and-thaw cycles [Fig.~\ref{fig:convergence}]. Additional
charge is localized in the $A$ cycles using the hybrid funcional in
the cluster (top row in Fig.~\ref{fig:convergence}). The density
spreads out again and the environment relaxes in the $B$ cycles when
the orbitals $B$ of the environment are optimized using the DFT
functional. However, the magnitude of these changes quickly decreases
with the iteration number and yields a well-converged density (and
well-converged energy) within 7 iterations.

The full hybrid and the converged embedded unpaired spin densities
closely match (Fig.~\ref{fig:methodcomp}) (b,c). By contrast, the DFT
density does not show a strong localization of the surplus electron at
all (Fig.~\ref{fig:methodcomp}) (a).  Indeed, projecting the converged
polaron orbital (i.e., the occupied majority spin Kohn-Sham orbital
with the highest energy) onto the central Ti atom of the distortion
yields quite small values for the overlap (0.39) for DFT, while the
full hybrid (0.69) and embedded calculations (0.65) agree quite
well. This underlines that despite the correct sign for the energy
gain when using the DFT orbitals in the hybrid energy functional, the
DFT description of the charge density is qualitatively deficient.

\section{Conclusions}

We have demonstrated a new embedding framework based on a suitable
rotation in the subspace of fully occupied Kohn-Sham orbitals. Using a
projection on local basis functions, a set of orbitals may be
localized at a site of interest, for example a defect. Subsequently,
these localized orbitals inside the cluster can now be optimized based
on a more expensive exchange-correlation functional, such as a hybrid
functional involving the exact evaluation of Fock exchange. Since
exchange interactions within the frozen environment are neglected,
the computation time is drastically reduced. The response of the
environment to the charge rearrangement in the cluster can be
self-consistently included by freeze-thaw cycles in which
alternatingly the orbitals in the embedded cluster or  in the
environment are optimized.

We have implemented our ansatz in the popular \texttt{VASP} software
package. As proof of principle, we have applied our method to two problems of
interest: a set of defects in bulk silicon, and small polarons in bulk
titania. We find excellent agreement with (much more expensive)
benchmark bulk hybrid calculations. 

\acknowledgements

The authors gratefully acknowledge support by the FWF via the SFB-41 ViCoM.

\end{document}